\documentclass[a4paper,11pt]{article}
\usepackage{jcappub} 
\usepackage{lineno}

\arxivnumber{1234.56789} 
\title{Measuring the curvature of the universe based on the dust-scattering ring of GRB 221009A}

\author{Jun-Yi Shen}
\author[1]{and Yuan-Chuan Zou\note{Corresponding author.}}
\affiliation{Department of Astronomy, School of Physics, Huazhong University of Science and Technology,
Wuhan 430074, China}

\emailAdd{zouyc@hust.edu.cn}

\abstract{
We investigate the use of dust-scattering rings from GRB 221009A to measure cosmic curvature. We derive the relationship between scattering angle and time delay in non-flat universes and attempt to constrain cosmological parameters by fitting theoretical predictions to observational data. The results show that this method is limited in its effectiveness due to the small geometric scale of observations. While unable to provide significant constraints, this study introduces a novel approach to cosmological measurements using GRB scattering rings. We discuss the method's limitations and potential future applications with larger-scale observations.
}

\begin{document}
\maketitle
\flushbottom

\section{Introduction}
\label{sec:intro}

Gamma Ray Bursts (GRBs) are an extremely high-energy burst phenomenon that occurs in the soft $\gamma$-ray wavelength.
GRBs has an isotropic equivalent luminosity reaching approximately $10^{53} \rm{\,erg \ s^{-1}}$ and originates from distant cosmological sources.
GRBs' extraordinary luminosity makes them some of the most distant observable phenomena in the universe. 
As $\gamma$-ray photons travel long cosmic distances, they carry valuable cosmological information, which can be utilized in cosmology studies. 
To obtain the luminosity distance from $E_{iso}$ and determine the corresponding redshift, which gives the cosmological parameter-determined luminosity distance, GRBs can serve as tools for constraining these cosmological parameters
\cite{2004ApJ...612L.101D,2021MNRAS.507..730H,2023ApJ...953...58L}. 
Similarly, GRBs can also be used to constrain the cosmology models \cite[e.g.][]{2015NewAR..67....1W,2021JCAP...09..042K}. 
Our work tries to use the photon's path of GRB 221009A to constrain the cosmological parameters. 

The Gamma-Ray Burst Monitor (GBM) onboard the Fermi satellite
was triggered on October 9th 2022 at 13:16:59.99 UT ($T_0$) on GRB 221009A \cite{2023ApJ...952L..42L}. 
This burst is the brightest event ever detected \cite{2023ApJ...946L..31B}. 
Following the trigger, other high-energy detectors also observed this event, such as Swift \cite{2023ApJ...946L..24W}. 
At $T_0+89$ ks, the luminosity of GRB 221009A significantly  decreased. The Swift XRT switched to photon counting mode and captured a two-dimensional image of the GRB's X-ray afterglow (see \cite{2023ApJ...946L..24W}, Fig. 3). 
A complex series of expanding dust-scattering echoes is clearly discernible within the image of GRB 221009A \cite{2023ApJ...946L..24W}. 
Before GRB 221009A, a few rings induced by the scattering of GRB photons had been identified, including those associated with GRB 061019, GRB 070129, and GRB 050724 ~\cite{2006ApJ...639..323V,2007A&A...473..423V}. 

The ring forms when a sudden burst of photons strikes a dust screen, scattering some of them.
Observers see a ring along the line of sight because the scattered photons at that angle arrive simultaneously. Over time, the ring expands.
Such an expanding ring can act as a tool to test the geometry of the universe, as the expanding speed is different for different (open/flat/closed) cosmology. 
Our work connects the geometry of the universe with GRB observations.
Before our work, observations of geometrical optics of gravitational Lensing by R$\ddot{a}$s$\ddot{a}$nen et al. were employed to test the Friedmann-Lemaître-Robertson-Walker (FLRW) metric and the spatial curvature parameter \cite{2015PhRvL.115j1301R}.
Our work also based on the geometrical optics scattered by dust.
However, the geometry is also degenerated with other parameters, such as $\Omega_{\rm M}$ and $\Omega_{\Lambda}$.
We try to use the rings of GRB 221009A to constrain the curvature of the universe. Meanwhile, as we apply our model, we find that the degenerate parameters have also been studied.  
One important is the expansion rate of the universe, a fundamental quantity in cosmological studies. 
The Hubble constant, denoted as $H_0$ relates the recession velocities of nearby sources with low redshift  $z$ to their distances. 
There are two distinct methods for measuring the Hubble constant. 
For two different way to measure the Hubble constant, the two measurements have divergence, see review ~\cite{2023arXiv231113305V}. This work should also constrain the $H_0$.

This paper is structured as follows. In section 2, we introduce the formation of scattering rings and explain how the recently detected rings of GRB 221009A can be processed  from data. In section 3, we connect the ring expansion to the cosmology model. In section 4, we show the fitting process of the cosmology parameters by the dust scattering rings observation of GRB 221009A. In section 5, we present the results. Finally, conclusion and discussion are given in section 6.

\begin{figure}[htbp]\label{fig:1}
\centering
\includegraphics[width=.5\textwidth]{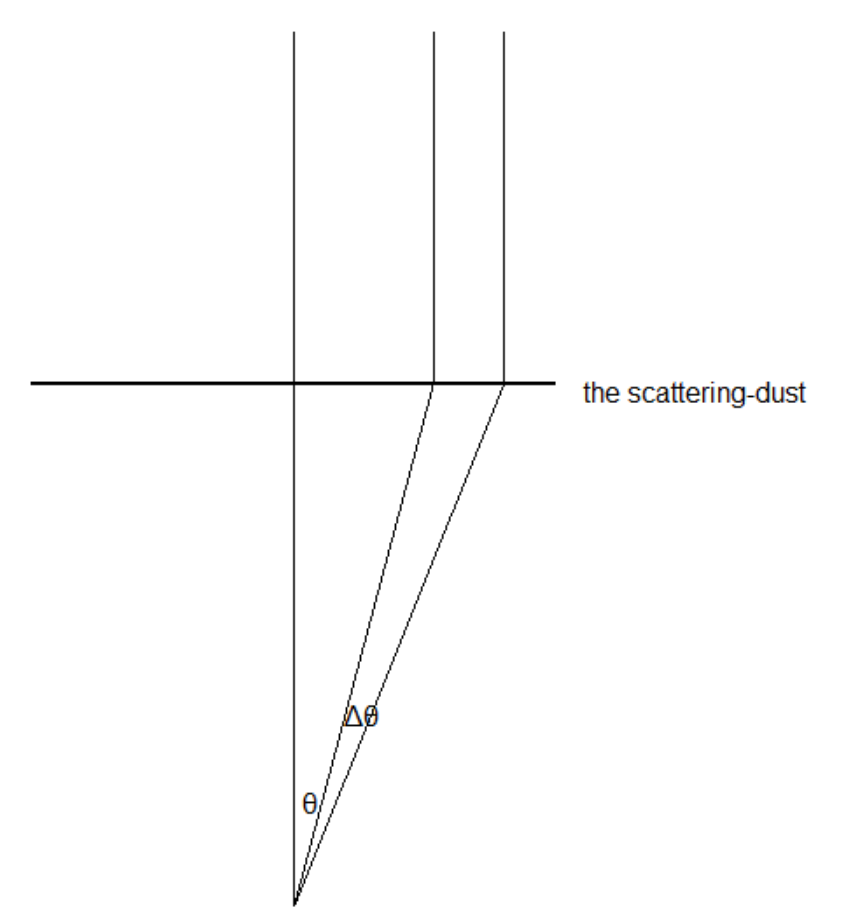}
\caption{Schematic plot for the dust scattering with edge-on view. Photons come from a distant source, with some of them being scattered by dust. The dust screen denotes by a thick line. We plot several optical paths. The non-scattering light path, which is a straight line, shows photons traveling this path arriving at the observer without delay. The scattered photons follow a longer path and therefore arrive later. 
$\Delta \theta$ corresponds to the expansion of the ring at a later time.
}
\end{figure}

\section{Echo rings of GRB }
Dust in galaxies interact with propagating photons in the ultraviolet or optical wavelengths, causing extinction and reddening. 
Concerning the $\gamma$-rays from a GRB, they exhibit particle-like behavior due to their short wavelength. 
This behavior results in the $\gamma$-rays being scattered by the dust pervading the galaxy. 
$\gamma$-rays produced by GRBs travel directly toward Earth from cosmological distances. 
Initially, photons travel long distances in parallel beams. Upon entering the Milky Way, they are partially scattered by dust particles, altering their direction. On Earth, an observer captures these nearly simultaneous photons in a single exposure lasting about 1,000 seconds. The scattered photons, detected by instruments, seem to originate from different directions than the source. As a result, the image brightens in multiple circular rings at angular radii from the center, where the source is located, while the multiple rings come from the multiple pulses of a certain GRB or multiple dust screens.
We plot a light path diagram to explain these rings, as shown in Fig. \ref{fig:1}. 
When the scattering angle is small ($\sim 10^{-4}$ rad), the delay time $\Delta t$ is related to the scattering angle $\theta$ as
\begin{equation}
    c \Delta t \approx \frac{ \theta^2 d }{2},
    \label{eq:ct}
\end{equation}
where $c$ is the speed of light and $d$ is the distance between dust in the galaxy and earth, typically on the order of  $\sim$ kpc.

After dividing the observation area of the sky into many pixels, non-scattered photons are detected at the source position in the sky
(for example, the source position of GRB 221009A is around RA $288.263^{\circ}$, DEC $19.803^{\circ}$).  
The scattered photons, arriving from other positions in the sky, were detected in adjacent pixels, allowing the scattering angle to be determined. 
The photons captured during a single exposure will produce a radial distribution in the image.
In certain pixels at specific radial positions, the photon counts will be denser, forming distinct rings.
These rings form because photons are scattered by dust at the same location and at the same angle
(corresponding to a delay time $\Delta t$, as shown in Eq. (\ref{eq:ct})), arrived simultaneously.

The rings caused by dust at the same location will expand over time.
The scattered photons, which have larger scattering angles and are scattered by the same dust as the later-arriving photons, will form larger rings.
By capturing images over a series of observational periods, we can analyze the growth rate of the rings.
This can be used to determine the position of the dust in the Milky Way as described in \cite{2023MNRAS.521.1590V}.

As described in \cite{2023ApJ...946L..24W}, Swift detected and reported the existence of scattered rings, including the dust scattering rings of GRB 221009A \footnote{\url{http:// www.swift.ac.uk/ user objects/ }}.
The data files containing information on the rings span Modified Julian Date (MJD) 59862–59866, with observation ID numbers
01126853004, 01126853005, 01126853006, 01126853008, and 01126853009. 
The rings can be found within these files under the corresponding IDs.
To find the rings in these data, first, utilizing the updated point-source function (PSF) for Swift \footnote{\url{https:// www.swift.ac.uk/ analysis/XRT/pileup.php }}, the radius profile of photons can be obtained. 
Using the Lorentzian function, the positions of rings with higher photon counts density in the radial distribution can be identified.
Vasilopoulos et al. processed the data \citep{2023MNRAS.521.1590V}. 
They utilized Markov chain Monte Carlo (MCMC) methods through the \texttt{emcee} package to determine the positions of the rings.
They determined $\theta$ correspond to $\Delta t$ of several series rings. In our work, we use the $\Delta t $ and $\theta$ values of rings found by them.

\section{Geometry of the photon's path} \label{sec:geo}

\begin{figure}[h]
    \centering
    \includegraphics[height=0.3\textwidth]{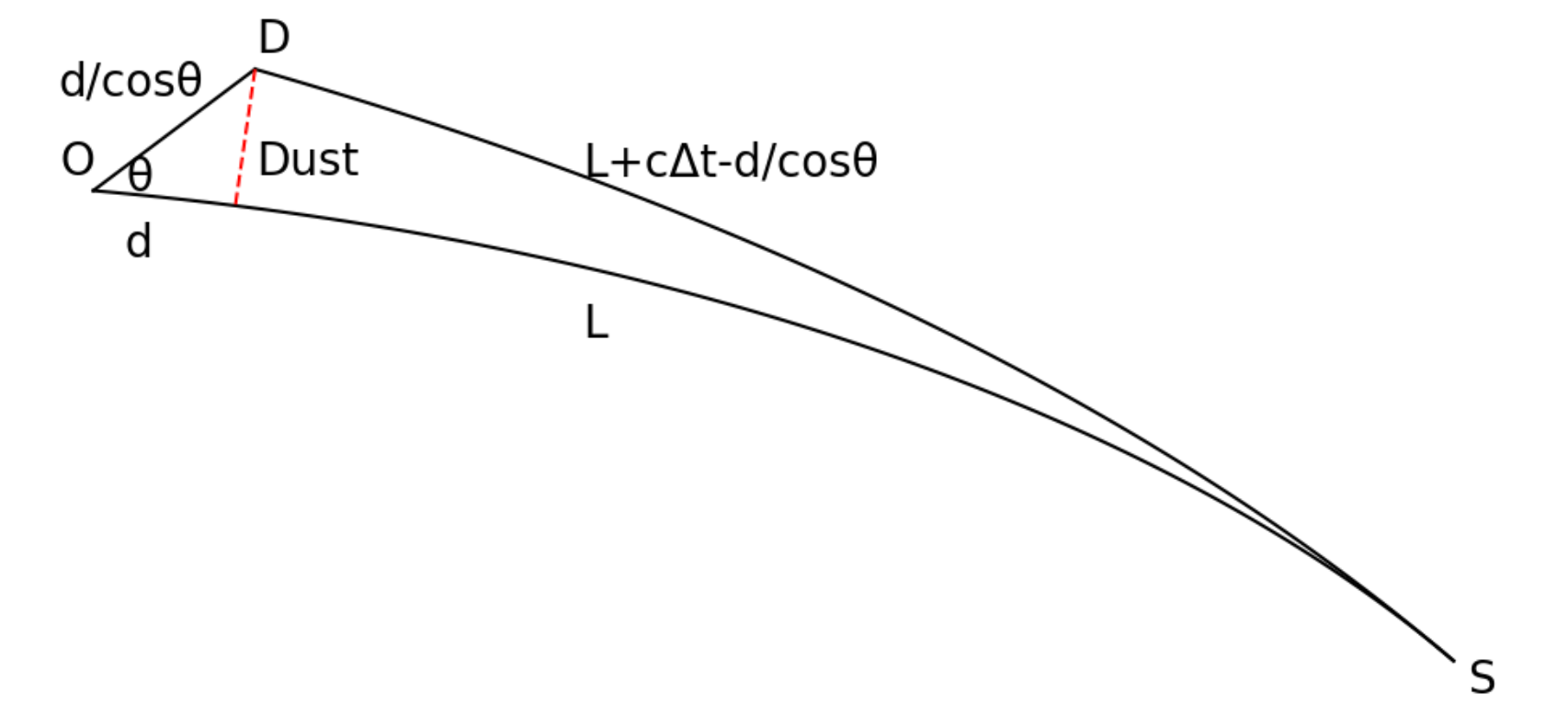}
    \caption{Schematic plot illustrating the formation of scattering-dust rings and the light paths in curved space. This figure assumes the curved space being a closed universe. 
    SO is the light path directly from the source S to the observer O. The dashed line crossing point D is the dust screen with edge-on view. SD and DO is the light path for the scattered light.
    Essentially, it is the same as Fig. \ref{fig:1} where SO and SD plotted as parallel lines, while we need a more precise representation in curved space rather than assuming SO and SD are parallel.
    }\label{fig:tri}
\end{figure}

Photons travel along geodesics. 
Considering two photons simultaneously originating from the source GRB 221009A, they follow two different trajectories as plotted in Fig. \ref{fig:tri}.
One trajectory represents non-scattered photons traveling directly to the observer, while the other represents photons scattered by galactic dust.
The photon traveling directly to Earth triggers the telescope at $T_0$, while the scattered photons, traveling a slightly longer distance, are observed at $T_0 + \Delta t$.
As shown in Fig. \ref{fig:tri}, we plot two trajectories: the paths SO (Source to Observer), SD (Source to Dust), and OD (Observer to Dust), all of which are geodesics.
These three points—observer O, source S, and dust D—form a geodesic triangle. 
SO represents the non-scattered trajectory, and SD plus OD represents the scattered trajectory. 
The scattered photons experience a delay of $\Delta t$. 
We use $L$ to denote the distance of non-scattered photons. 
Since the path SD + OD has a $\Delta t$ delay, the length of SD equals to $L+c\Delta t-d/\mathrm{cos\theta}$. 
For a flat universe with $\Omega_{\rm{k}}=0$, this triangle adheres to the cosine rule in flat spacetime obeys the cosine theorem, i.e., 

\begin{equation}\label{eq:falt}
    (L+c\Delta t-d/\mathrm{cos\theta})^2=L^2+(d/\mathrm{cos\theta})^2-2Ld.
\end{equation}

In Eq. (\ref{eq:falt}), the quantities $L, \theta, \Delta t$ are determined through  observation.
In flat spacetime, Eq. (\ref{eq:falt}) allows us to derive $\theta $ as a function of $\Delta t(\theta)$.
From this equation, the only parameter $d$ can be estimated, as described in \cite{2023MNRAS.521.1590V}. 
They used this to study several dust distribution.
However, for a non-flat universe, the cosine law in non-Euclidean geometry must be reformulated as follows:
\begin{equation}
    \mathrm{cosn}\frac{(L+c\Delta t-d/\mathrm{cos\theta})}{\rho}=\mathrm{cosn}\frac{L}{\rho}\mathrm{cosn}\frac{d}{\mathrm{cos\theta} \rho} \pm \mathrm{sinn}\frac{L}{\rho}\mathrm{sinn}\frac{d}{\mathrm{cos\theta} \rho} \mathrm{cos \theta}.\label{eq:cos}
\end{equation}
Here, we unify the closed and open universe into a single formula. 
The function $\rm{cosn}$ represents the hyperbolic cosine for $\Omega_{\rm{k}}>0$, corresponding to an open universe, and the cosine for $\Omega_{\rm{k}}<0$, corresponding to a closed universe. 
The sign ``$\pm$" indicates that ``+" corresponds to an open universe, while ``–"corresponds to a closed universe.
$\rho $ denotes the curvature radius of the universe. Both $L$ and $\rho$ change over time due to the universe's expansion. 
We consider their present-day values. 
$\Omega_{\rm{k}}$ is a cosmological parameter indicating the curvature of the universe.
An important survey by the Planck collaboration indicates $\Omega_{\rm{k}} = 0.001 \pm 0.002$, corresponding to $\rho \sim 100$ Gpc \cite{2020A&A...641A...6P}.
From the Eq. (\ref{eq:cos}), we can infer that the expansion speeds of the rings vary slightly across different types of universes. 
However, the differences are minimal, as elaborated in the following description. 
Eq. (\ref{eq:cos}) can be simplified due to several small quantities: $c \Delta t \sim 10^{16}$ cm $\ll d \sim \rm{kpc} \ll L|_{(z \sim 0.1)}  \sim  $ Gpc. 
We can expand Eq. (\ref{eq:cos}) in a Taylor series, considering terms of the order of $O(d/\rho)^2$. 
By relating these terms and treating $d/\rho$ and $c \Delta t/\rho$ as expandable quantities, we obtain:
\begin{equation}
      \frac{d \theta^2}{1 - \frac{ d }{ L \mathrm{cos \theta} } \mathrm{cotn}(\frac{L}{\rho})\frac{L}{\rho}}=2 c \Delta t. \label{eq:reWcos}
\end{equation}
With this equation, $\theta$ can be obtained for a given $\Delta t$. 
When compared to flat spacetime, Eq. (\ref{eq:reWcos}) reduces to the flat case:
\begin{equation}
    d \theta^2\left(1 + \frac{ d }{ L \mathrm{cos \theta} }\right) =2 c \Delta t \label{eq:recos}.
\end{equation}
As suggested in observations like Planck collaboration observations \cite{2020A&A...641A...6P}, $|\Omega_{\rm{k}}| \ll 1$.
Therefore, the Eq. (\ref{eq:reWcos}) can be further simplified under the condition $L \ll \rho $, i.e.,
\begin{equation}
      d \theta^2 \left\{ 1 + \frac{ d }{ L \mathrm{cos \theta} } \left[1+\frac{1}{3}\left(\frac{L}{\rho}\right)^2\right] \right\} =2 c \Delta t. \label{eq:fin}
\end{equation}

\section{Parameter fitting}
According to \cite{1998AJ....116.1009R}, they selected the quantity distance modulus $\mu$ for cosmological parameters' simulation. 
The distance modulus $\mu$ can be obtained from supernova Ia observations or theoretically inferred.
Similarly, we choose the observable quantity $\theta$ of GRB scattering rings, which can also be theoretically determined.
Next, we describe how we fit the cosmological parameters under the most widely accepted cosmological model, the $\Lambda$CDM model. 
This model posits that the universe is isotropic and homogeneous on large scales, described by the FLRW metric. 
In the FLRW metric, distances are determined by redshift $z$ and three cosmological parameters: the mass density $\Omega_{\rm{M}}$, vacuum energy density $\Omega_{\Lambda}$, and Hubble constant $H_0$.
The co-moving distance is
\begin{equation}
    D_{\rm c} =  cH_0^{-1} |\Omega_{\rm{k}}|^{-1/2}  \mathrm{sinn}  {|\Omega_{\rm{k}}|^{1/2}  \times 
    \int_0^z {\rm d}z[(1+z)^2(1+\Omega_{\rm{M}} z)-z(2+z)\Omega_{\Lambda}]^{-1/2}
    },
    \label{eq:D_c}
\end{equation}
where $\Omega_{\rm{k}}+\Omega_{\Lambda}+\Omega_{\rm{M}}=1$.
The distance from the GRB source, $L$, can be written as $L(z,\Omega_{\rm{k}},\Omega_{\rm{M}},H_0)$, which can be derived from Eq. (\ref{eq:D_c}). 
Then, Eq. (\ref{eq:reWcos}) can be written as (considering $z$ as a quantity determined by observation, for each event $z$ can be regarded as constant): 
\begin{equation}
    \theta = \left[2 c \Delta t \left(1 + \frac{ d }{ \rho(\Omega_{\rm{k}}, H_0) }\mathrm{cosn\frac{ \emph{L}(\Omega_{\rm{k}},\Omega_{\rm{M}},\emph{H}_0) }{\rho(\Omega_{\rm{k}}, \emph{H}_0)}}\right)\right]^{1/2}, \label{eq:thth}
\end{equation}
In the $\Lambda$CDM model, $\rho$ is defined as:
\begin{equation}
    \rho = \frac{c}{ H_0 \sqrt{|\Omega_{\rm{k}}|}}.
    \label{eq:rho}
\end{equation}
The subscript  ``0'' denotes quantities in the present day.
We approximate $\rm{cos\theta} $ as $ \sim 1$ in Eq. (\ref{eq:reWcos}).
Eq. (\ref{eq:thth}) provides a series of theoretical values of ${\theta}(\Omega_{\rm{k}}, \Omega_{\rm{M}}, H_0 ,\Delta t)$. 
There are N number of observed rings, and also N theoretical $\theta$s. We use the least squares algorithm to fit the parameters.
Following the approach in supernova cosmology as in \cite{1998AJ....116.1009R}, we write a $\chi^2$:
\begin{equation}
    \chi^2=\sum_i \frac{ [\theta_i(\Omega_{\rm{k}}, \Omega_{\rm{M}}, H_0 ,\Delta t_i)-\theta_{obs,i}]^2}{\sigma_{i}^2},
\end{equation}
where the subscript $i$  represents different exposure moment with various scattering echo. The $\sigma$ denotes the errors of observation.
We used the theoretical $\theta$ to simulate the cosmology constant.
Next, we consider the possibility distribution function (PDF) of the cosmology parameters. 
The PDF can be derived from Bayes theorem ($\boldsymbol {\theta}$ denotes a set of $N$ theoretical values of $\theta$s):
\begin{equation}
    p(\Omega_{\rm{k}},\Omega_{\rm{M}},H_0 \ | \boldsymbol{\theta} )= \frac{p(\Omega_{\rm{k}},\Omega_{\rm{M}},H_0)p( \boldsymbol {\theta} | \Omega_{\rm{k}},\Omega_{\rm{M}},H_0)}{p( \boldsymbol {\theta})}.
\end{equation}
Since we have no prior constrains of PDF $p(\boldsymbol {\theta}), p(\Omega_{\rm{k}},\Omega_{\rm{M}},H_0)$, we take them as constant. Then, we have :
\begin{equation}
    p(\Omega_{\rm{k}},\Omega_{\rm{M}},H_0 \ | \boldsymbol{\theta} ) \propto p(\boldsymbol {\theta} | \Omega_{\rm{k}},\Omega_{\rm{M}},H_0)
\end{equation}
If the $\theta$ obeys  independent and Gaussian distribution, we have: 
\begin{equation}\label{eq:sim}
    p(\boldsymbol {\theta} | \Omega_{\rm{k}},\Omega_{\rm{M}},H_0) = \Pi_i \frac{1}{\sqrt{2\pi \sigma_i^2}} \times \mathrm{exp}  \left\{\frac{[\theta_i( \Omega_{\rm{k}},\Omega_{\rm{M}},H_0, \Delta t_i)- \theta_{obs}]^2}{2 \sigma_i^2}\right\}.
\end{equation}
Therefore, we have:
\begin{equation} 
    p(\boldsymbol {\theta} | \Omega_{\rm{k}},\Omega_{\rm{M}},H_0) \propto \mathrm{exp}(\frac{- \chi ^2}{2}).
\end{equation}
The minimized $\chi^2$ corresponds to the maximum possibility of parameters.

\section{Result} \label{sec:result}
The host galaxy of the GRB 221009A is at a redshift of approximately $ z \sim 0.151$.
From Eq. (\ref{eq:D_c}), we have $L = \rm{D_c}(z|_{(0.151)}, H_0, \Omega_{\rm{M}},\Omega_{\Lambda})$. 
Besides the cosmological parameters, the fitted equation still has two unknown quantities, $d$ and $\Delta t$. $d$ is the distance between the dust and the observer. 
We use the measurement of  $d$ in Minkowski spacetime from \cite{2023MNRAS.521.1590V}, where d is fixed to be $ \sim 2 $ kpc.
The $\Delta t$ we use roughly corresponds to the observing time minus $T_0$. 
Actually, $\Delta t$ should represent a time interval between $T_0$ and the moment of exposure, with an associated error bar.
The error shall be a uniformly distributed. 
This approach is not very rigorous, but it does not impact our result here since we do not require a highly precise $\Delta t$. 
Therefore, in the fitting formula, only the cosmology parameters affect $\chi^2$.
The observational values $d$ and $\Delta t$ in this fitting were given by  \cite{2023MNRAS.521.1590V}.
They used Lorentz function fitting to determine the echo ring positions $\theta$ in each exposure dataset. 
Their results are presented in Table \ref{tab:params}. 
Using MCMC, we minimized the  $\chi^2$ in attempt to fit the cosmological parameters. 
Unfortunately, the results of the MCMC were not satisfactory, as the steps did not converge. 
We investigated this issue and found that the minimization area could be narrow. 
We also attempted to use the Monte Carlo (MC) method to minimize  $\chi^2$. 
We show our MC results in Table \ref{tab:2}, where we use different parameter sets: P1 [$H_0$=$70 \ \rm{km \ s^{-1} \ Mpc^{-1} }$, $\Omega_{\rm{k}}=$ 1, $\Omega_{\rm{M}}=$ 1], P2 [$70 \ \rm{km \ s^{-1} \ Mpc^{-1} }$, 0.1, 0.1], P3 [$70\ \rm{km \ s^{-1} \ Mpc^{-1} }$, 0.5, 0.5] to calculate their $\chi^2$ for reference.

\begin{table*}
\caption{ The position of rings $\theta$ corresponds to the arrival time $T_0+\delta t$ of the scattering rings of GRB 221009A. We partially list three rings, $r_1,r_2 $ and $r_3$, each scattered by different dust. The data were fitted by Vasilopoulos et al. \cite{2023MNRAS.521.1590V}, and we utilized their fitted data. The unit of position is arcseconds.}
\label{tab:params}
\centering
\begin{tabular}{*{11}{c}}
\hline 
rings ID  & \multicolumn{5}{c}{MJD}\\
  & 
 59862.66 &
 59862.79 &
 59862.87 &
 59863.05 &
 59863.26 &
  \\ 
\hline 

$r_{1}$ &2.466$_{-0.018}^{+0.017}$ &  2.590$_{-0.026}^{+0.026}$ &2.578$_{-0.012}^{+0.006}$   &2.838$_{-0.028}^{+0.029}$ & 3.116$_{-0.021}^{+0.022}$ & \\

$r_{2}$ &3.342$_{-0.04}^{+0.029}$ & 3.56$_{-0.05}^{+0.05}$ &  3.65$_{-0.05}^{+0.04}$  &3.91$_{-0.07}^{+0.06}$ & 4.18$_{-0.05}^{+0.05}$ &  \\

$r_{3}$ &5.700$_{-0.009}^{+0.009}$ & 5.971$_{-0.017}^{+0.019}$ & 6.242$_{-0.016}^{+0.016}$  &6.629$_{-0.016}^{+0.016}$ & 7.115$_{-0.013}^{+0.014}$ & \\
   
\hline 
\end{tabular}

\end{table*}

\begin{table}[h]
\caption{Our best fitting gives the $\chi^2_m$. We also show the $\chi^2$ of three different parameters P1, P2, P3.  This table shows that $\chi^2$ is not sensitive to these parameters.}

    \label{tab:2}
    \centering
    \begin{tabular}{c|c}
        \hline
        parameters & $\chi^2/\chi^2_m$ \\
        \hline
        P1 &  1+2e-3 \\
        \hline
        P2 &  1+3e-3 \\
        \hline
        P3 &  1+2e-3 \\
    \end{tabular}
\end{table}

Such a result means any values of cosmological parameters (obey the requirement assumed in section \ref{sec:geo}) fit the observation equally well.
Therefore, the current observed rings cannot give a good constraint for the cosmological parameters yet. 
The reason can be seen in Eq. (\ref{eq:fin}). The cosmological curvature only shows in the term where $\rho$ appears. However, a small number $d/L$ should always be multiplied in that term. This makes the exact value of $\rho$ is not essential. 
If $d/L$ is not negligible or $\theta$ is highly precise, $\rho$ can be determined in principle.

Now, we discuss the conditions under which we can confirm the cosmological parameters based on our approach. 
We consider the most favorable condition. 
It is evident that larger observational geometries provide better outcomes for constraining cosmological curvature. 
For a good outcome, $L$ cannot be much less than $\rho$ (not a local, small scale). Therefore, the scattering positions should not be within the galaxy.
If $ d \ll L$, the fitting formula remains unchanged and is still given by Eq. (\ref{eq:reWcos}). 
This condition is also excluded. 
Therefore, we assume that the GRB is scattered by a galaxy along its path. 
The distance between the scattering point and the observer, $d/\cos\theta$, is not local and cannot be regarded as Minkowski spacetime. Therefore, $d/\cos\theta$ should be rewritten as:
\begin{equation}
    d'=\mathrm{arctann}(\mathrm{tann}(\frac{d}{\rho})/\mathrm{cos}\theta).
\end{equation}
Under this condition, the relation between $\rho$ and $\theta$ is given by:
\begin{equation}
    \mathrm{cosn}\frac{(L+c\Delta t-d')}{\rho}=\mathrm{cosn}\frac{L}{\rho}\mathrm{cosn}\frac{d'}{ \rho} \pm \mathrm{sinn}\frac{L}{\rho}\mathrm{sinn}\frac{d'}{ \rho} \mathrm{cos \theta}.\label{eqnoex}
\end{equation}

\begin{figure}
    \centering
    \includegraphics[width=0.7\textwidth]{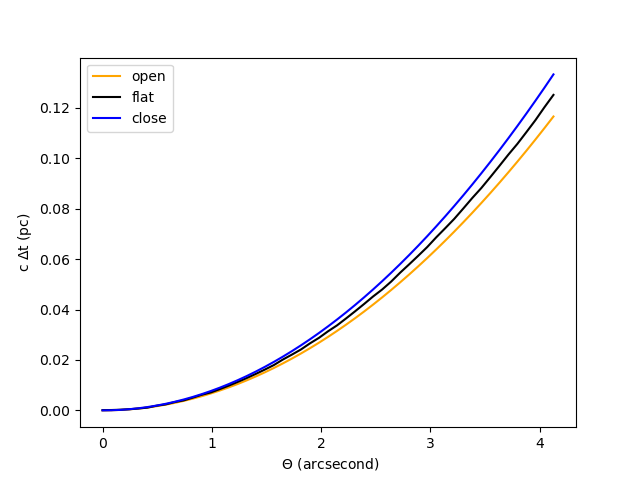}
    \caption{ For our most favorable condition, $L=10$ Gpc, $d=2$ Gpc, the curves representing the open and closed conditions with $\rho$ is 10 Gpc equal to L as solved based on the relevant equations \ref{eqnoex}. $\theta$ is close to the observation revolution, like Swift UVOT $\sim 1''$. The time delay $\Delta t$ is approximately 3 months, which is plausible for observation. However, even under these most favorable conditions, we still cannot obtain a good constraint for $\rho$. }
    \label{fig:ideal}
\end{figure}

The maximum $L$ corresponds to a redshift $z \sim 1$, which is approximately 10 Gpc. 
A relatively favorable resolution for $\rho$ is $\sim 10L$. 
If $\rho > 10L$, we cannot easily distinguish between different parameter curves. 
Given the observation conditions, such as the resolution of the {\it Swift}/UVOT
\footnote{\url{https://swift.gsfc.nasa.gov/about_swift/uvot_desc.html }}, the delay time is reasonable. 
As shown in Fig. \ref{fig:ideal}, our best fit for $\rho (\Omega_{\rm{k}})$ indicates $\rho>10$ Gpc or $\Omega_{\rm{k}}<0.1$, which is less accurate compared to other methods for measuring $\Omega_{\rm{k}}$. 
From this figure, one can see that if sub-arcsecond measurement of the ring is achievable, the geometry can in principle be identified.  
The future observations on the rings from cosmological distant galaxy might be a good probe to measure the flatness of the universe.

\section{Conclusion and Discussion}
GRB 221009A is an extremely luminous and rare GRB event that provides valuable insights into the nature of GRBs.
The geometry of the GRB photons' trajectory, coupled with the curvature of the universe, is reflected in the dust scattering rings.
We investigate how these rings can help us study the universe through these photons.
However, current data is still not precise enough to make good constraint on the geometry.
We also tried using the MC method to find suitable parameters. 
However, despite varying the formula, we did not achieve satisfactory results. 
To minimize $\chi^2$, we randomly scattered points in the parameter space and calculated their corresponding $\chi^2$ values, but no significant improvement was observed.
The reason could be understood as following. 
Dividing Eq. (\ref{eq:D_c}) by Eq. (\ref{eq:rho}), we find that the magnitude of $L/\rho$ is not strongly connected with the parameters $\Omega_{\rm{M}}, \Omega_\Lambda$, and $\Omega_{\rm{k}}$. 
We denote the $L/\rho \equiv x$. 
The function ${\rm cotn} x \cdot x$ in parameter fitting Eq. (\ref{eq:reWcos}) changes slowly. 
The independent variable $x$ in this equation ranges from $0$ to $1$. 
Only when $x$ increases significantly, ${\rm cotn} x \cdot x$ changes a few times. 
Therefore, even the parameters $H_0, \Omega_{\rm{k}}, \Omega_{\rm{M}}$ vary widely, the differences in $\chi^2$ are minimal.
Consequently, we could not get a stringent constraint. 

Though the current dust rings from GRB 221009A cannot determine the geometry of the universe precisely yet, the method is novel and could be applied to future observations. As discussed in section \ref{sec:result}, future ring observation from distant galaxy ($\sim 1$ Gpc) might be able to distinguish the geometry, which requires a bursting event passing through a foreground galaxy with a dust screen. 
There are still several challenges. 
1. Such a configuration requires the host galaxy of the source, the host galaxy of the dust ring and the observer lies in the same line. The chance is significantly lower than the configuration that the dust ring located inside the Milky Way. More other transients such as flares of blazars, tidal disruption events, should be considered to enhance the chance. 
2. Weak gravitational Lensing should play a role in the light path. One may carefully degenerate this from the geometry of the universe. 
3. Resolving the angular diameter of the rings from a distant galaxy requires more powerful X-ray telescopes. Considering the burst observation requires large field of view (FOV), future telescopes such as Athena (FOV greater than 40 arcmin, angular resolution 3 arcsec)\citep{2013arXiv1306.2307N} and PFA onboard eXTP (FOV greater than 12 arcmin, angular resolution 15 arcsec) \citep{2019SCPMA..6229502Z} may play an important role. 


\acknowledgments
This work is supported by the National Key R\&D Program of China (2021YFC2203100).
The computation was completed on the HPC Platform of Huazhong University of Science and Technology.


\bibliographystyle{JHEP}
\bibliography{biblio.bib}

\providecommand{\href}[2]{#2}\begingroup\raggedright\begin{thebibliography}{10}

\bibitem{2004ApJ...612L.101D}
Z.G.~{Dai}, E.W.~{Liang} and D.~{Xu}, \emph{{Constraining {\ensuremath{\Omega}}$_{M}$ and Dark Energy with Gamma-Ray Bursts}}, \href{https://doi.org/10.1086/424694}{\emph{\apjl} {\bfseries 612} (2004) L101} [\href{https://arxiv.org/abs/astro-ph/0407497}{{\ttfamily astro-ph/0407497}}].

\bibitem{2021MNRAS.507..730H}
J.P.~{Hu}, F.Y.~{Wang} and Z.G.~{Dai}, \emph{{Measuring cosmological parameters with a luminosity-time correlation of gamma-ray bursts}}, \href{https://doi.org/10.1093/mnras/stab2180}{\emph{\mnras} {\bfseries 507} (2021) 730} [\href{https://arxiv.org/abs/2107.12718}{{\ttfamily 2107.12718}}].

\bibitem{2023ApJ...953...58L}
J.-L.~{Li}, Y.-P.~{Yang}, S.-X.~{Yi}, J.-P.~{Hu}, F.-Y.~{Wang} and Y.-K.~{Qu}, \emph{{Constraints on the Cosmological Parameters with Three-Parameter Correlation of Gamma-Ray Bursts}}, \href{https://doi.org/10.3847/1538-4357/ace107}{\emph{\apj} {\bfseries 953} (2023) 58} [\href{https://arxiv.org/abs/2306.12840}{{\ttfamily 2306.12840}}].

\bibitem{2015NewAR..67....1W}
F.Y.~{Wang}, Z.G.~{Dai} and E.W.~{Liang}, \emph{{Gamma-ray burst cosmology}}, \href{https://doi.org/10.1016/j.newar.2015.03.001}{\emph{\nar} {\bfseries 67} (2015) 1} [\href{https://arxiv.org/abs/1504.00735}{{\ttfamily 1504.00735}}].

\bibitem{2021JCAP...09..042K}
N.~{Khadka}, O.~{Luongo}, M.~{Muccino} and B.~{Ratra}, \emph{{Do gamma-ray burst measurements provide a useful test of cosmological models?}}, \href{https://doi.org/10.1088/1475-7516/2021/09/042}{\emph{\jcap} {\bfseries 2021} (2021) 042} [\href{https://arxiv.org/abs/2105.12692}{{\ttfamily 2105.12692}}].

\bibitem{2023ApJ...952L..42L}
S.~{Lesage}, P.~{Veres}, M.S.~{Briggs}, A.~{Goldstein}, D.~{Kocevski}, E.~{Burns} et~al., \emph{{Fermi-GBM Discovery of GRB 221009A: An Extraordinarily Bright GRB from Onset to Afterglow}}, \href{https://doi.org/10.3847/2041-8213/ace5b4}{\emph{\apjl} {\bfseries 952} (2023) L42} [\href{https://arxiv.org/abs/2303.14172}{{\ttfamily 2303.14172}}].

\bibitem{2023ApJ...946L..31B}
E.~{Burns}, D.~{Svinkin}, E.~{Fenimore}, D.A.~{Kann}, J.F.~{Ag{\"u}{\'\i} Fern{\'a}ndez}, D.~{Frederiks} et~al., \emph{{GRB 221009A: The Boat}}, \href{https://doi.org/10.3847/2041-8213/acc39c}{\emph{\apjl} {\bfseries 946} (2023) L31} [\href{https://arxiv.org/abs/2302.14037}{{\ttfamily 2302.14037}}].

\bibitem{2023ApJ...946L..24W}
M.A.~{Williams}, J.A.~{Kennea}, S.~{Dichiara}, K.~{Kobayashi}, W.B.~{Iwakiri}, A.P.~{Beardmore} et~al., \emph{{GRB 221009A: Discovery of an Exceptionally Rare Nearby and Energetic Gamma-Ray Burst}}, \href{https://doi.org/10.3847/2041-8213/acbcd1}{\emph{\apjl} {\bfseries 946} (2023) L24} [\href{https://arxiv.org/abs/2302.03642}{{\ttfamily 2302.03642}}].

\bibitem{2006ApJ...639..323V}
S.~{Vaughan}, R.~{Willingale}, P.~{Romano}, J.P.~{Osborne}, M.R.~{Goad}, A.P.~{Beardmore} et~al., \emph{{The Dust-scattered X-Ray Halo around Swift GRB 050724}}, \href{https://doi.org/10.1086/499353}{\emph{\apj} {\bfseries 639} (2006) 323} [\href{https://arxiv.org/abs/astro-ph/0511351}{{\ttfamily astro-ph/0511351}}].

\bibitem{2007A&A...473..423V}
G.~{Vianello}, A.~{Tiengo} and S.~{Mereghetti}, \emph{{Dust-scattered X-ray halos around two Swift gamma-ray bursts: GRB 061019 and GRB 070129}}, \href{https://doi.org/10.1051/0004-6361:20077968}{\emph{\aap} {\bfseries 473} (2007) 423} [\href{https://arxiv.org/abs/0707.2343}{{\ttfamily 0707.2343}}].

\bibitem{2015PhRvL.115j1301R}
S.~{R{\"a}s{\"a}nen}, K.~{Bolejko} and A.~{Finoguenov}, \emph{{New Test of the Friedmann-Lema{\^\i}tre-Robertson-Walker Metric Using the Distance Sum Rule}}, \href{https://doi.org/10.1103/PhysRevLett.115.101301}{\emph{\prl} {\bfseries 115} (2015) 101301} [\href{https://arxiv.org/abs/1412.4976}{{\ttfamily 1412.4976}}].

\bibitem{2023arXiv231113305V}
L.~{Verde}, N.~{Sch{\"o}neberg} and H.~{Gil-Mar{\'\i}n}, \emph{{A tale of many $H_0$}}, \href{https://doi.org/10.48550/arXiv.2311.13305}{\emph{arXiv e-prints} (2023) arXiv:2311.13305} [\href{https://arxiv.org/abs/2311.13305}{{\ttfamily 2311.13305}}].

\bibitem{2023MNRAS.521.1590V}
G.~{Vasilopoulos}, D.~{Karavola}, S.I.~{Stathopoulos} and M.~{Petropoulou}, \emph{{Dust-scattering rings of GRB 221009A as seen by the Neil Gehrels Swift X-ray Observatory: can we count them all?}}, \href{https://doi.org/10.1093/mnras/stad375}{\emph{\mnras} {\bfseries 521} (2023) 1590} [\href{https://arxiv.org/abs/2302.02383}{{\ttfamily 2302.02383}}].

\bibitem{2020A&A...641A...6P}
{Planck Collaboration}, N.~{Aghanim}, Y.~{Akrami}, M.~{Ashdown}, J.~{Aumont}, C.~{Baccigalupi} et~al., \emph{{Planck 2018 results. VI. Cosmological parameters}}, \href{https://doi.org/10.1051/0004-6361/201833910}{\emph{\aap} {\bfseries 641} (2020) A6} [\href{https://arxiv.org/abs/1807.06209}{{\ttfamily 1807.06209}}].

\bibitem{1998AJ....116.1009R}
A.G.~{Riess}, A.V.~{Filippenko}, P.~{Challis}, A.~{Clocchiatti}, A.~{Diercks}, P.M.~{Garnavich} et~al., \emph{{Observational Evidence from Supernovae for an Accelerating Universe and a Cosmological Constant}}, \href{https://doi.org/10.1086/300499}{\emph{\aj} {\bfseries 116} (1998) 1009} [\href{https://arxiv.org/abs/astro-ph/9805201}{{\ttfamily astro-ph/9805201}}].

\bibitem{2013arXiv1306.2307N}
K.~{Nandra}, D.~{Barret}, X.~{Barcons}, A.~{Fabian}, J.-W.~{den Herder}, L.~{Piro} et~al., \emph{{The Hot and Energetic Universe: A White Paper presenting the science theme motivating the Athena+ mission}}, \href{https://doi.org/10.48550/arXiv.1306.2307}{\emph{arXiv e-prints} (2013) arXiv:1306.2307} [\href{https://arxiv.org/abs/1306.2307}{{\ttfamily 1306.2307}}].

\bibitem{2019SCPMA..6229502Z}
S.~{Zhang}, A.~{Santangelo}, M.~{Feroci}, Y.~{Xu}, F.~{Lu}, Y.~{Chen} et~al., \emph{{The enhanced X-ray Timing and Polarimetry mission{\textemdash}eXTP}}, \href{https://doi.org/10.1007/s11433-018-9309-2}{\emph{Science China Physics, Mechanics, and Astronomy} {\bfseries 62} (2019) 29502} [\href{https://arxiv.org/abs/1812.04020}{{\ttfamily 1812.04020}}].

\end{thebibliography}\endgroup


\end{document}